\documentstyle[11pt,aaspp4]{article}

\received{}
\accepted{}

\lefthead{Buzasi et al.}
\righthead{Seismology of $\alpha$ UMa}

\begin{document}

\title{The Detection of Multimodal Oscillations on $\alpha$ UMa}

\author{D. Buzasi}
\affil{Space Sciences Laboratory, University of California, 
Berkeley, CA 94720}

\author{J. Catanzarite, R. Laher, T. Conrow, D. Shupe}
\affil{Infrared Processing and Analysis Center, California Institute of
Technology, MS 100-22,
Pasadena, CA 91125}

\author{T. N. Gautier III}
\affil{Jet Propulsion Laboratory, California Institute of Technology, MS 100-22,
Pasadena, CA 91125}

\author{T. Kreidl}
\affil{Information Technology Services and 
Department of Physics and Astronomy, Box 5100, Northern Arizona University, 
Flagstaff, AZ 86011}

\and

\author{D. Everett}
\affil{NASA/Goddard Space Flight Center, Greenbelt, MD 20771}



\begin{abstract}
We have used the star camera on the WIRE satellite to observe the K0~III
star $\alpha$ UMa, and we report the apparent detection of 10 oscillation modes. The
lowest frequency mode is at $1.82~\mu \rm Hz$, and appears to be the fundamental
mode. The mean spacing between the mode frequencies is $2.94~\mu \rm Hz$,
which implies that all detected modes are radial. The mode frequencies are
consistent with the physical parameters of a K0~III star, if we assume that
only radial modes are excited. Mode amplitudes are 
$100 - 400~\mu \rm mag$, which is consistent with the scaling relation of
Kjeldsen \& Bedding \markcite{kandb95}(1995).
\end{abstract}


\section{Introduction}

Over the past few decades, our understanding of the interior of the Sun -- its
thermodynamic structure, internal rotation, and dynamics -- has been revolutionized
by the technique of helioseismology, the study of the frequencies and amplitudes
of seismic waves that penetrate deep into the solar interior (Leibacher et al. 
\markcite{leibacher85}1985;
Duvall et al. \markcite{duvall88}1988; 
Schou et al. \markcite{schou98}1998). The high quality of the modeling 
in the solar case is made possible by the large number ($\approx 10^7$) of
modes visible in the Sun. Unfortunately, the lack of spatial resolution 
inherent in stellar observations limits the number of detectable modes in stars
to only a few (those with low degree $l$). Nonetheless, successful detection of
even a few modes has the potential to provide greatly improved values for
fundamental stellar parameters such as mass, abundance, and 
age (Gough \markcite{gough87}1987; Brown et al. \markcite{brown94}1994)

While oscillations have been
successfully detected on roAp stars, $\delta$
Scuti stars, and white dwarfs, these stars all show oscillation amplitudes
several orders of magnitude larger than is expected from solar-like stars.
More recently, however, a number of authors 
(Hatzes \& Cochran \markcite{hatzes94a}1994a, \markcite{hatzes94b}1994b;
Edmonds \& Gilliland \markcite{edmonds96}1996) have reported the
detection of periodic variability at levels of a few hundred $\rm m~s^{-1}$
and/or several millimagnitudes in a number of K giants. However, each
detection is of only a single mode,
and multimodal oscillations have yet to be unambiguously detected in any 
cool star other than the Sun.

In March 1999, NASA launched the Wide-Field Infrared Explorer satellite, with
the intent of carrying out an infrared sky survey to better understand 
galaxy evolution.
Unfortunately, within days of launch the primary science instrument on WIRE failed
due to loss of coolant. However, the satellite itself continues to function
nearly perfectly, and in May we began a program of asteroseismology using the 
WIRE's onboard 52 mm aperture star camera. Below we report WIRE's probable detection
of multimodal oscillations on a cool star, which is the first of its kind.

\section{Instrument Description and Data Reduction Technique}

The WIRE satellite star camera, a Ball Aerospace model CT-601, 
consists of a $512 \times 512$ SITe CCD with 
$27 \mu$m pixels (1 arc minute on the sky) and gain of $15 \rm~e^-/ADU$, 
fed by a 52 mm, f/1.75 refractive optical system. The read noise of the system
is 30 electrons. The pixel
data is digitized with a 16 bit ADC, and up to five $8 \times 8$ pixel fields
can be digitized and transmitted to the ground. For this work, we used only
one field, which permitted us to read out the CCD at a rate of 10 Hz. The
stellar image is somewhat defocused, but essentially all of the light falls
on the central $2 \times 2$ pixel spot. The spectral response of the system is 
governed entirely by the response of the CCD plus the optical system, and
is approximately equivalent to the V+R bandpass.

The WIRE satellite is in a sun-synchronous orbit which, when combined with
constraints imposed by the solar panels, limits pointing to two strips, each approximately
$\pm 30^\circ$ wide,
located perpendicular to the Earth-Sun line.
In addition, continuous observing is not possible. Early in the program, we
were able to obtain only 7 or 8 minutes worth of data during every 96-minute
orbit. Later, after viewing constraints were relaxed (which involved
scheduling software and onboard data-table modifications), observing efficiency
rose to as much as 40 minutes per target per orbit -- up to two targets
are possible during any orbit. Thus, with integrations
every 0.1 s, we acquired continuous data segments of up to 24,000 observations.

Bias correction was performed on board the satellite, and further 
data reduction was accomplished using software developed at IPAC.
Each $8 \times 8$ pixel 
field was extracted by summing the flux in the central $4 \times 4$ pixel
region. Although scattered light in the field is limited by the one-meter
sun-shield mounted on the star camera, we performed a background
subtraction using the flux from a 20-pixel octagonal annulus
surrounding the central region
of the image. Finally, we converted our fluxes to an instrumental magnitude.
After removal of thermal effects (see below), the rms noise in the final 
reduced time series was typically comparable to
the 1.8 mmag noise expected from pure photon statistics, although non-Poisson
noise is certainly present as well.
The lack of a good
flat field for the instrument was a concern, which we dealt with by
rejecting those frames in which the mean centroid of the stellar image lay
more than $4\sigma$ from the mean position, where $\sigma$ is the mean 
standard deviation of the image centroid. This criterion applied to approximately
3.8\% of the observations, and the vast majority of these observations were at the
start or end of an orbital segment.
Overall, the satellite displayed excellent attitude stability during our run, 
with $\sigma$
measured to be typically 0.7 arc sec or less (Laher et al. 2000).

\section{Observations and Data Analysis}

$\alpha$ UMa was the primary target for WIRE from 18 May through 23 June 1999.
It is a K0 III star (Taylor \markcite{taylor99}1999)
with an angular diameter of
$6.79 \rm~mas$ (Hall \markcite{hall96}1996; 
Bell \markcite{bell93}1993). At the Hipparcos distance of 38 pc,
this corresponds to a stellar radius of $28 \rm~R_{\sun}$, in substantial
agreement with the earlier value of $25 \rm~R_{\sun}$ derived by Bell (1993).
The effective temperature of the star is variously reported as 
$T_e = 3970 \rm~K$ to $T_e = 4660 \rm~K$ (Cayrel de Stroebel \markcite{cayrel92}
1992), with the
latter value being the most recent (Taylor \markcite{taylor99}1999).

$\alpha$ UMa is a member of a binary system, with a total system mass of
$5.94 \rm~M_{\sun}$ (Soderhjelm \markcite{soderhjelm99}1999). 
The spectral type of $\alpha$ UMa B
is somewhat unclear, with F7~V most often cited in the literature (see, e.g.,
the SAO catalog). However, on the basis of IUE observations, Kondo, Morgan,
\& Modisette \markcite{kondo77}(1977) estimate the 
secondary to be ``late A'' (see also
Ayres, Marstad, \& Linsky \markcite{ayres81}1981). 
In either case, however, the secondary makes
only a small contribution ($\approx 6\%$) to the total system luminosity, and
should show oscillation frequencies quite different from those of the primary.

During the observation period, WIRE made a
total of $4,036,448$ observations of $\alpha$ UMa, after removal of those observations with
poor pointing characteristics.
The data were first phased at the period of the spacecraft, to determine if
any obvious instrumental periodicities existed. This phasing showed the existence
of a strong sinusoidal variation, with amplitude $\approx 8$ mmag. A thermistor
is mounted on the star camera, and examination of data from this thermistor 
showed that these variations were in fact correlated with temperature variations
of a few tenths of a degree. Although a thermoelectric cooler (TEC) is mounted
on the CCD, the star tracker thermal design lowered the CCD temperature below the
default setpoint, so the TEC never actually turned on.
In order to reduce the impact of this signal
on our data analysis, we prewhitened the data by fitting and subtracting from the
entire time series a sinusoid constrained
to have the satellite orbital period (the phase was allowed to vary). The best-fit
sinusoid has an amplitude of 8.63 mmag, and its subtraction results in an rms residual
of 2.05 mmag, compared to the 1.84 mmag expected from photon statistics.
We also explored other means of fitting and removing the thermal signature,
including high-order polynomial fits to the phased data. Such approaches, 
while more complex than sine fitting, did not lead to any appreciable
improvement in the fit, and were therefore discarded. The use of
any of these thermal fitting procedures did not affect the peaks in the amplitude
spectrum described below. In fact, these peaks were visible even before 
the application of any fit to the thermal variation, though removal of the 
thermal signature, by removing the largest single peak, did help to render them more easily visible.

Data from the first portion of the run (18 May - 7 June) were all in short
($t < 8 \rm~minutes$) segments, which obviously extended over a short range of
orbital phase ($\approx 0.08$). We were concerned that including these data in our
sine fit would bias the result, so we decided to exclude them from our analysis. In addition, 
NASA responded to our expressed concern about the TEC by lowering the set
point on 18 June, and the CCD behavior subsequent to this had not yet 
reached an equilibrium point before the end of our run. Thus, we excluded
data taken after 18 June from the analysis as well, leaving only the 8 June
-- 18 June window of approximately $9.2 \times 10^5 \rm~s$. Shortening the
time series clearly adversely affects our frequency resolution, but we felt
that confidence in the data quality was the overriding issue. The data that
were subjected to analysis are shown in Figure 1, after removal of the 
thermal variation. 

Data were searched for periodicities using Discrete Fourier Transform (DFT;
Foster \markcite{foster96}1996),
Lomb-Scargle periodogram (Scargle \markcite{scargle82}1982; 
Horne \& Baliunas \markcite{horne86}1986), 
and epoch-folding techniques (Davies \markcite{davies90}1990)
which are essentially equivalent to phase dispersion modulation
(PDM; see Stellingwerf \markcite{stellingwerf78}1978; 
Schwarzenberg-Czerny \markcite{schwarzenberg98}1998).  
The Scargle periodogram analysis was conducted as described in Scargle (1982;
see also Hatzes \& Cochrane 1994ab); the data were windowed using a Parzen function
and the resulting periodogram was oversampled
by a factor of 8 in frequency space. The DFT analysis was similarly oversampled,
and implemented the CLEAN algorithm of Roberts, 
Lehar, \& Dreher \markcite{roberts87}(1987) to
remove alias peaks. When using the DFT, the data were windowed using a 
Parzen function, and 100
iterations of CLEAN were performed. Due to its relative slowness, epoch-folding
analysis was conducted only within the frequency range of interest, as 
identified by the Scargle and DFT analysis, and was used only to aid
in interpretation of the
results from the other two algorithms.
In the discussion that follows, we will concentrate on the
Lomb-Scargle periodogram results, though, in general, the three techniques
gave similar results. 

Figure 2 shows the window function for the time series. The upper frequency limit
for the figure has been arbitrarily set at $5~\rm mHz$ to enhance the visibility of the
amplitude spectrum at frequencies near 1 mHz, while the lowest frequencies are shown
in the inset. No significant features are present in the window function above
5 mHz. The large evenly spaced peaks correspond to the satellite orbital
frequency and its aliases. 

The Lomb-Scargle periodogram
for the time series is shown in Figure 3 on the same frequency scale as
the window function. 
The low-frequency inset shows ten significant peaks, 
and the frequencies and amplitudes of these peaks are given in 
Table~1 (derived from Lorentzian fits),
along with conservative formal error estimates derived from the half-width 
of the 
periodogram peaks. We note that periodograms of portions
of the time series (halves and thirds)
give results similar to that of the whole, with decreased frequency resolution,
and that sine fits to these frequencies show coherent phasing in these different
portions.

\placetable{table-1}

Unfortunately, the implementation of the on-board data collection on WIRE means that we lack 
simultaneous
observations of a comparison star, and are thus essentially performing absolute photometry
with an instrument not designed for that purpose. However, we have 
observed stars
other than $\alpha$ UMa, and the sun-synchronous orbit of WIRE implies that most
instrumental effects should be similar for all sources.
The dashed line in the Figure~3 inset shows the periodogram from a time series
of $\alpha \rm~Leo$, a B7~V star not expected to show significant low-frequency
oscillations. The $\alpha$ Leo data set, which was obtained from 23 May through
3 June 1999, consists of segments similar in length to those of $\alpha$ UMa, has similar
rms noise to the $\alpha$ UMa time series, and was reduced in exactly the
same manner. Our object here is not to perform analysis of the $\alpha$
Leo data (which might well benefit from a different approach than we have
used for $\alpha$ UMa), but rather to show that the particular low frequency
peaks in the periodogram of $\alpha$ UMa do not arise from either
instrumental effects or the data reduction procedure itself. The larger
peaks in the $\alpha$ Leo periodogram may arise from an imperfect
removal of both long-term and orbital variations; none show coherent
phasing across different segments of the time series.
The dissimilarity of the two periodograms increases our confidence
that the peaks
visible in the $\alpha$ UMa periodogram are due to the star itself,
although it is of course possible that the instrumental 
behavior changes significantly for different targets.

As is apparent from Figures 2 and 3, the family of peaks
visible at low frequencies is repeated at higher frequencies, which leads to the difficulty
of determining which set of peaks is the correct one. We can easily eliminate peaks above
$\approx 200 ~\mu \rm Hz$ by examining the summed power spectrum of the individual orbital
segments, which shows no significant power at these frequencies. We are therefore left with
the problem of selecting between the set of peaks in the $\approx 2 - 50 ~\mu \rm Hz$ range
and the similar set around $200 ~\mu \rm Hz$. We believe that the low-frequency peaks
are the physical solution and the higher-frequency set an alias for the following
two reasons:

\begin{enumerate}

\item The low-frequency peaks are always of larger amplitude than the alias 
peaks. While a resonance of stochastic noise with an alias frequency can enhance
an individual alias peak such that it is larger than the corresponding true
peak, this is
unlikely to occur simultaneously for multiple peaks. 
We have performed simulations which
confirm this reasoning. 

\item{Hatzes (1999, private communication) has searched for oscillations in
$\alpha$ UMa using ground-based spectroscopic methods. He reports finding
frequencies of 1.36 and 6.0 $\mu \rm Hz$ (the latter less convincingly), 
although his observing run was too
short to lend much confidence to the exact values. While not identical to the
frequencies we report here, these frequencies are certainly comparable to
our lower-frequency peaks rather than to the alias peaks.} 

\end{enumerate}

Though neither of these factors is conclusive on its own, we believe that
together they indicate that we are on solid ground interpreting the
observed peaks in the amplitude spectrum as stellar oscillations. Of course, it 
remains possible that the observed variations are due to instrumental effects
or to non-oscillatory stellar phenomena such as granulation.

\section{Interpretation}

Below we discuss the astrophysical implications of our results. 
A more detailed discussion, in the context
of a complete stellar interiors model for $\alpha$ UMa A, can be found
in Guenther et al. \markcite{guenther99}(1999).

\subsection{Mode Frequencies and Spacings}

The frequency of the fundamental mode is determined primarily by structure in
the envelope and so its determination requires a complete stellar model. However,
we can easily determine the range in which it should lie. The fundamental 
period $P_0$ is given 
by (Christy \markcite{christy66}1966; \markcite{christy68}1968)
\begin{equation}
Q_0 = P_0 \sqrt{\frac{\rho}{\rho_{\sun}}}
\end{equation}
where $Q_0$ should lie between the value of 0.038 for a polytrope with $\gamma
= 4/3$ and 0.116 (for $\gamma = 5/3$). Using the values of $R = 28 R_{\sun}$
from interferometry and the mass $M \approx 4 M_{\sun}$ appropriate to a
K0 giant yields a fundamental mode of between 2.8 and 8.6 days. The lowest
frequency that we see corresponds to a period of 6.35 days, so we identify it
as the fundamental mode for $\alpha$ UMa.

As noted above, the average mode spacing for the first 8 modes is 
$2.94 ~\mu \rm Hz$. The last two modes have much larger spacings, which we interpret
as signifying that we are not detecting all of the possible oscillation modes
for the star, presumably because they are not excited to large amplitudes.
The large separation $\Delta \nu_0$ is related to the mean stellar density, 
as shown by Cox \markcite{cox80}(1980):
\begin{equation}
\Delta \nu_0 = 135 \left( \frac{\rho}{\rho_{\sun}} \right)^{1/2} ~\mu \rm Hz
\end{equation}
Using the values appropriate to a typical K0 giant yields a predicted spacing of 
$1.82 ~\mu \rm Hz$, about half the observed value. Once again, this 
discrepancy can be accounted for by assuming that all modes are not
excited. In particular, if only even or odd-valued radial $n$ modes are 
excited, we
would expect the observed large separation to be twice the predicted value.
The simplest explanation is then that only the $l=0$ modes are excited, and
the frequencies we observe correspond to radial oscillations of the star.

\subsection{Mode Amplitudes}

The Kjeldsen \& Bedding \markcite{kandb95}(1995) scaling law
\begin{equation}
\delta L / L = \frac{L/L_{\sun}}{(\lambda/550 {\rm~nm})(T_e/5777{\rm~K})^2
(M/M_{\sun})} \times 5.1 \mu \rm mag
\end{equation}
predicts oscillation amplitudes of $\approx 500 ~\mu \rm mag$ for $\alpha$
UMa, which is essentially in agreement with our results. It should be noted
that the WIRE data are obtained in white light and, consequently, phase differences
in the oscillation amplitudes as a function of wavelength would tend to combine to
reduce the observed amplitude. Consequently, it is not surprising that the
observed amplitudes are somewhat smaller than those predicted by theory.
In addition, of course, extending a relationship derived for lower main sequence
stars to giants is a risky enterprise!
Nonetheless, the near-agreement between theory and observation may imply that the
excitation mechanism for oscillations in $\alpha$ UMa is fundamentally 
similar to the solar mechanism (presumably convection; see, e.g. Bogdan
et al \markcite{bogdan93}1993), unlike oscillations observed in other K giants,
which show amplitudes an order of magnitude greater than those we have 
detected.

\acknowledgments

We gratefully acknowledge the support of Dr. Harley Thronson and Dr. Phillipe Crane
at NASA Headquarters for making this unusual use of WIRE possible.
J.C., T.C., R.L., T.G., and D.S. would like to thank Drs. Carol Lonsdale and Perry Hacking
for the opportunity to work with them on the WIRE project,
and their support of the WIRE asteroseismology effort. The hard work of many
people, including the WIRE operations and spacecraft teams at GSFC, and
the timeline generation team
team at IPAC, was essential to making this project a reality.
We would also like to acknowledge the contributions of the anonymous referee,
whose criticisms helped to greatly improve the presentation of our results.


\clearpage

\figcaption[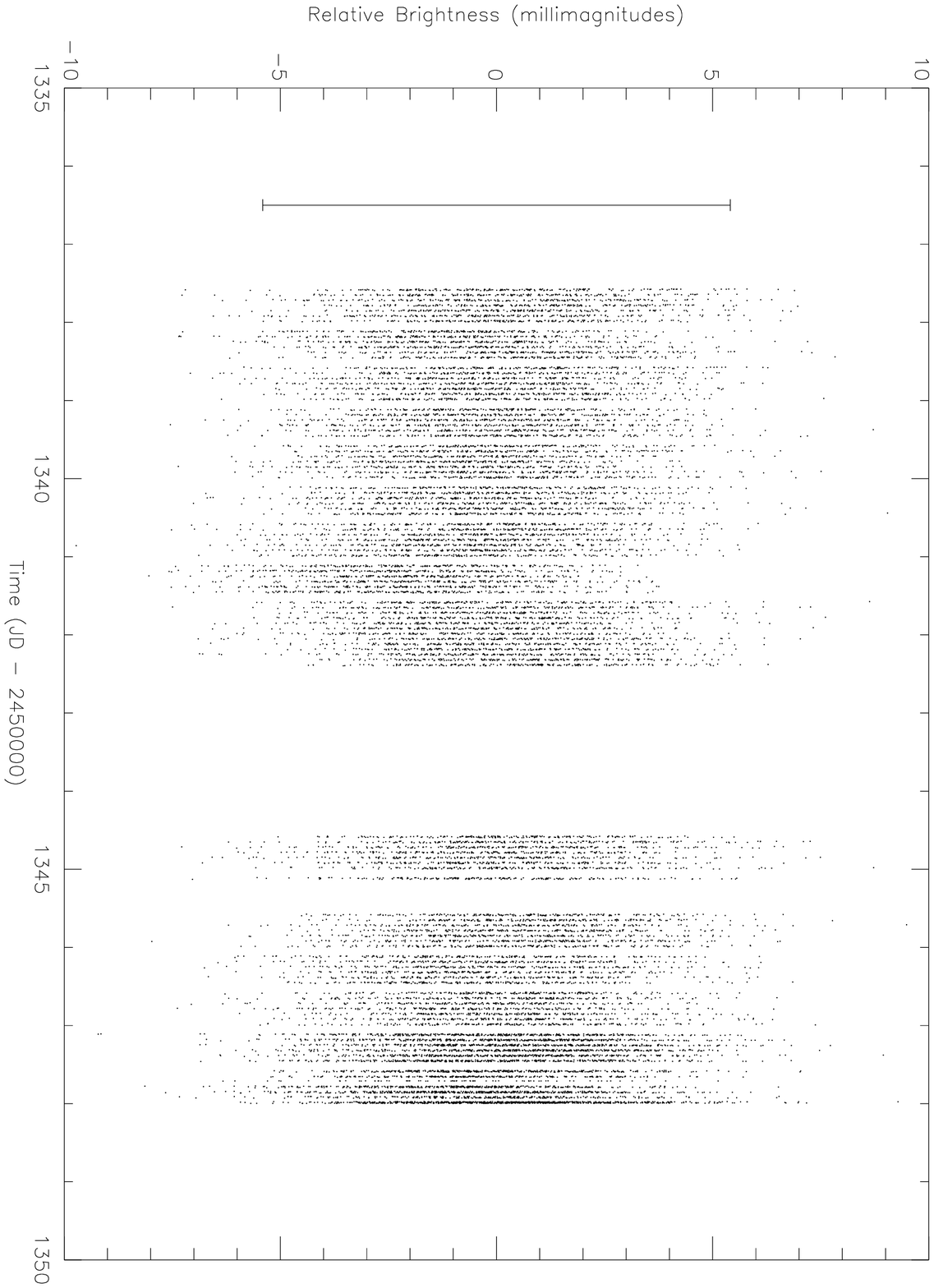]{The thermally corrected and mean subtracted time series 
of observations of $\alpha$ UMa.
Each vertical stripe visible in the figure corresponds to one spacecraft orbit
(about 96 minutes). The error bar shown represents the $\pm 3 \sigma$ 
Poisson noise
based on pure photon statistics. \label{fig1}}

\figcaption[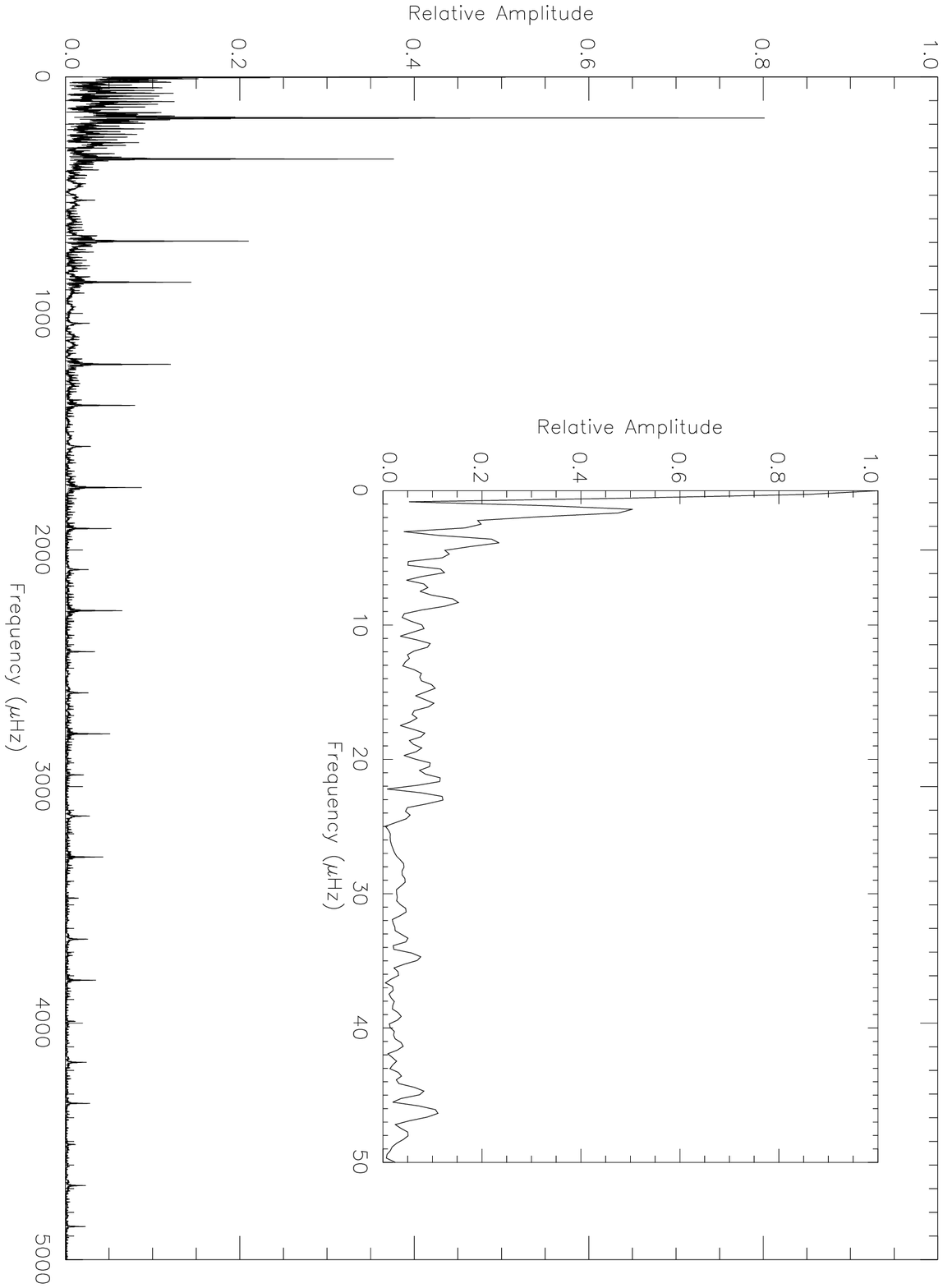]{The window function for the $\alpha$ UMa time series
shown in Figure~1. \label{fig2}}

\figcaption[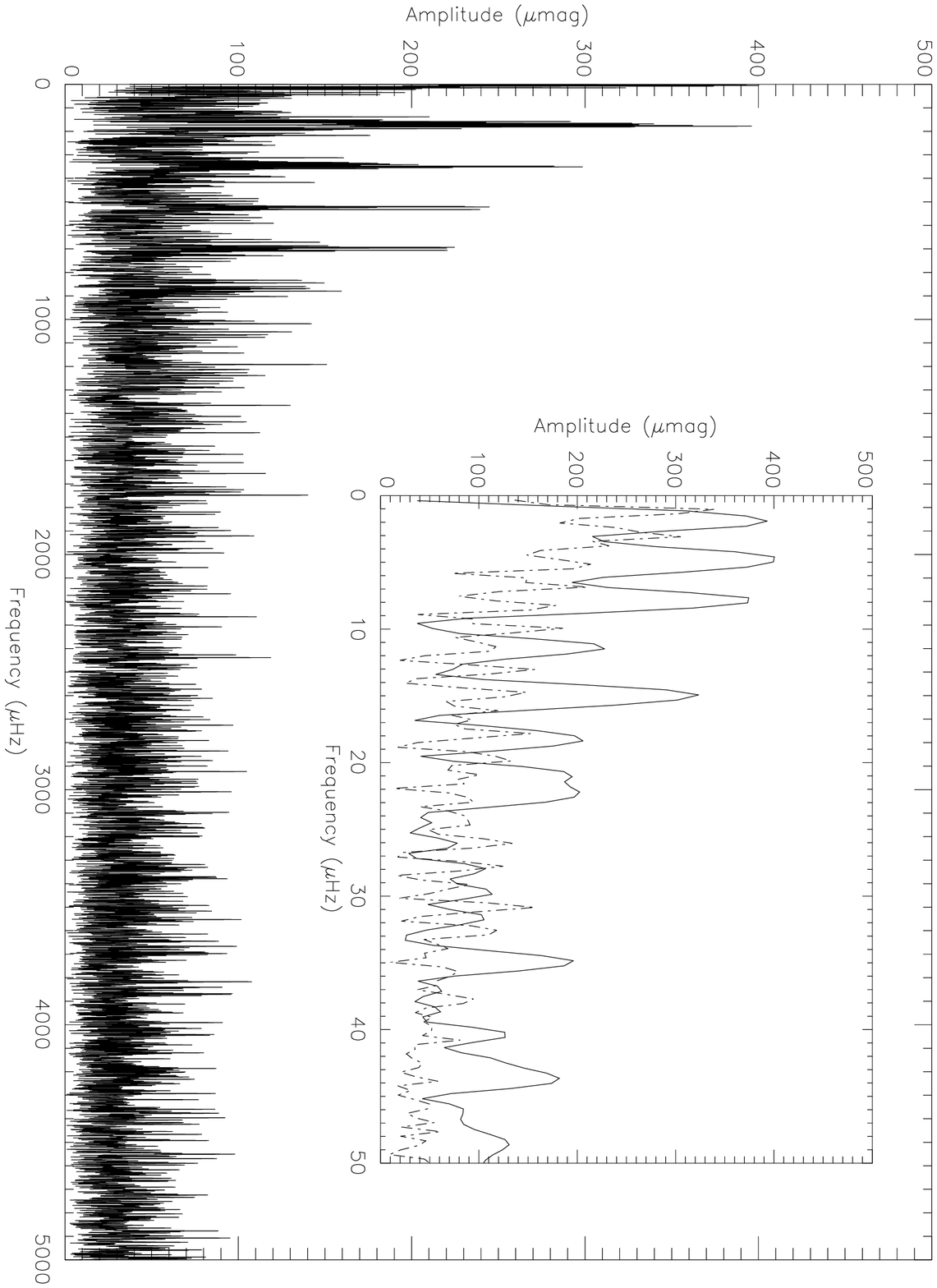]{The Scargle periodogram of the time series in 
Figure~1. The dashed line in the inset represents the periodogram of a
comparable time series of $\alpha$ Leo. \label{fig3}}

\clearpage

\plotone{figure1.eps}

\clearpage

\plotone{figure2.eps}

\clearpage

\plotone{figure3.eps}

\clearpage 

\begin{table*}
\begin{center}
\begin{tabular}{cccc}
Frequency ($\mu\rm Hz$)& Error ($\mu\rm Hz$) & $\Delta \nu$ 
($\mu\rm Hz$)\tablenotemark{1} & Amplitude ($\mu\rm mag$)\\
\tableline
1.82 & 0.74\tablenotemark{2}& & 390 \\
4.84 & 0.98 & 3.02 & 400 \\
7.86 & 0.77 & 3.02 & 390 \\
11.41 & 0.59 & 3.55 & 230 \\
15.00 & 0.71 & 3.59 & 330 \\
18.25 & 0.71 & 3.25 & 220 \\
20.90 & 0.67 & 2.65 & 200 \\
22.37 & 0.73 & 1.47 & 210 \\
34.93 & 0.63 & 12.56 & 210 \\
43.56 & 0.92 & 8.63 & 180 \\
\end{tabular}
\end{center}

\tablenotetext{1}{The difference in frequency between each peak and the previous
one. The mean separation for the first 8 modes is 2.94 $\mu$Hz.}
\tablenotetext{2}{Since this frequency is poorly sampled by the time series, the 
error estimate for the lowest
frequency is probably an underestimate
of the true error, which may be as much as twice this value.}
\tablenum{1}
\end{table*}

\end{document}